\documentclass{aastex631}

\usepackage{graphicx}	
\usepackage{amsmath}	
\usepackage{amssymb}	
\usepackage[utf8]{inputenc}
\usepackage[T1]{fontenc}
\usepackage[english]{babel}
\usepackage{soul}
\usepackage{mhchem}

\shorttitle{SiS formation in the interstellar medium}
\shortauthors{Mota et al.}
\graphicspath{{./}{figures/}}

\begin{document}

\title{SiS formation in the interstellar medium through Si+SH gas phase reactions}

\author[0000-0001-8368-0803]{V. C. Mota}
\affiliation{Departamento de Física, Universidade Federal do Espírito Santo, 29075-910 Vitória, Brazil}

\author[0000-0003-1501-3317]{A. J. C. Varandas}
\affiliation{Departamento de Física, Universidade Federal do Espírito Santo, 29075-910 Vitória, Brazil}
\affiliation{School of Physics and Physical Engineering, Qufu Normal University, 273165, PR China}
\affiliation{Coimbra Chemistry Centre and Chemistry Department, University of Coimbra, 3004-535 Coimbra, Portugal}

\author[0000-0001-9381-7826]{E. Mendoza}
\affiliation{Observatório do Valongo, Universidade Federal do Rio de Janeiro, Ladeira Pedro Antônio, 43, Rio de Janeiro – RJ 20.080-090, Brazil}

\author[0000-0001-9676-2605]{V. Wakelam}
\affiliation{Laboratoire d’astrophysique de Bordeaux, Univ. Bordeaux, CNRS, B18N, allée Geoffroy Saint-Hilaire, 33615 Pessac, France; \href{mailto:valentine.wakelam@u-bordeaux.fr}{valentine.wakelam@u-bordeaux.fr} }

\author[0000-0002-4184-2437]{B. R. L. Galvão}
\affiliation{Centro Federal de Educação Tecnológica de Minas Gerais, CEFET-MG, Av. Amazonas 5253, 30421-169, Belo Horizonte, Minas Gerais, Brazil;\href{mailto:brenogalvao@gmail.com}{brenogalvao@gmail.com} }




\begin{abstract}
Silicon monosulfide is an important silicon bearing molecule detected in circumstellar envelopes and star forming regions.
Its formation and destruction routes are not well understood, partially due to the lack of a detailed knowledge on the
involved reactions and their rate coefficients.
In this work we have calculated and modeled the potential energy surface (PES) of the HSiS system employing highly 
accurate multireference electronic structure methods. After obtaining an accurate analytic representation of the PES,
which includes long-range energy terms in a realistic way via the DMBE method, we have calculated 
rate coefficients for the Si+SH$\rightarrow$SiS+H reaction over the temperature range of 25-1000K. 
This reaction is predicted to be fast, with a rate coefficient of $\sim 1\times 10^{-10}\rm cm^3\, s^{-1}$ at 200K, which substantially increases for lower temperatures 
(the temperature dependence can be described by a modified Arrhenius equation with  $\alpha=0.770\times 10^{-10}\rm cm^3\,s^{-1}$, $\beta=-0.756$ and $\gamma=9.873\, \rm K$).
An astrochemical gas-grain model of a shock region similar to  L1157-B1 shows that the inclusion of the Si+SH reaction 
increases the SiS gas-phase abundance relative to \ce{H2} from $5\times 10^{-10}$ to $1.4\times 10^{-8}$, which perfectly matches  the observed abundance of $\sim 2\times 10^{-8}$.
\end{abstract}



\section{\label{sec:intro}Introduction}

Silicon monosulfide is the simplest molecule to contain a silicon-sulphur chemical bond and may play an important role in the formation of sulfide grains in the interstellar medium (ISM)~\citep{ZHU08:229,MAS19:A62}. After its first detecion in the envelope of an evolved carbon star~\citep{morris1975detection}, it has been observed in several other environments~\citep{dickinson1981interstellar,danilovich2018sulphur,lefloch2018astrochemical} together with its isovalent counterpart, SiO. From these observations, SiS is consistently found to be less abundant than SiO, and although the origin of silicon monoxide is { satisfactorily understood}, this is not the case for SiS.
Evidence for a strong gradient of [SiO/SiS] ratio across a shock region has been 
obtained~\citep{podio2017silicon}, indicating different chemical pathways. 
As a result of a systematic search for SiS in low-mass star forming regions \citep{lefloch2018astrochemical} 
it was concluded that SiS cannot be a simple product of gas phase chemistry in molecular 
clouds, but rather requires specific conditions, which are  best found in shocks.
In circumstellar envelopes, on the other hand, SiS is believed to be formed in the densest and hottest parts of the inner envelope, 
where it is thought to be formed under thermochemical conditions \citep{prieto2015si,schoier2007abundance}.

Given the importance of the SiS molecule as a potential gas-phase precursor to sulfide grains, its unknown chemistry and poorly explained abundance, several neutral-neutral
formation and destruction reactions have been explored recently, both theoretically and experimentally. For example, \cite{zanchet2018formation} proposed that one of the main SiS formation reactions in outflows is Si+SO. 
Further theoretical calculations showed that Si+SH, SiH+S and SiH+S$_2$ collisions \citep{ROS18:695,ROS19:306,PAI20:299} are also potential candidates.
As for destruction routes, it has been computationally shown that SiS is stable towards collisions with both atomic and molecular hydrogen~\citep{PAI20:299}, but is efficiently destroyed in reaction with atomic oxygen~\citep{paiva2018accurate,zanchet2018formation}.
Recent laboratory studies~\citep{DOD21:7003} together with astrochemical models have also highlighted the importance of the reaction Si+\ce{SH2 -> SiS + H2} in star forming regions, in spite of its nonadiabatic character.

Recent astrochemical models predict that the Si+OH reaction is one of the major sources of interstellar SiO in the gas phase~\citep{GUS08:695,SAN17:1675}.
It should also be interesting to verify whether the analogous Si+SH reaction could also be a major source of SiS, but accurate values for its rate coefficients are necessary for this verification.  For this reason, we develop in this work an accurate potential energy surface for the HSiS triatomic system, which allows for the calculation of accurate rate coefficients for SiS formation via S+SiH and Si+SH reactions, both of which are exothermic and barrierless. 
Quasiclassical trejectories are employed to obtain accurate rate constants for the title reaction, which are employed in an astrochemical model to study its role in shock regions.

\section{Theoretical methods}

\subsection{Electronic structure calculations}
\label{sec:2}


The product channel SiS($^1\Sigma^+$)+H($^2S$) correlates with the PES of a single electronic state of 
the HSiS molecule, namely $^2A'$. In turn, the $^2A''$ and quartet states correlate with the excited 
SiS($a^3\Pi$) molecule~\citep{PAI20:299}, which lies much higher in energy and does not contribute to formation of SiS. For this reason, only the ground $^2A'$ PES has been calculated and modeled in this work.

The calculations have been performed with the MOLPRO package~\citep{MOLPRO} employing the internally contracted version of the MRCI~\citep{WER88:5803} method { including single and  double excitations and
the Davidson correction (MRCI+Q). The full valence CASSCF wave function~\citep{WER85:5053} was used as reference. 
Specifically, the 1s, 2s and 2p orbitals of Si and S were considered as core, and the remaining 9 orbitals of the system were treated as active with 11 electrons (11e/9o).} 
A large number of single point energies covering the whole configuration space of the molecule were calculated with the aug-cc-pV$(Q+d)$Z and aug-cc-pV$(5+d)$Z basis set~\citep{DUN89:1007,KEN92:6796}, which have been subsequently extrapolated to the complete basis set \citep{VAR18:177} (CBS) limit.

The CASSCF energies were extrapolated with the physically motivated exponential scheme~\citep{FEL93:7059,VAR19:8022}
\begin{equation}
E_X^{HF}=E_{\infty}^{HF}+B\exp{(-\beta X),}  
\label{eq:CBSHF}
\end{equation}
with $B$ and $E_{\infty}^{HF}$ being parameters to be determined from a fit to a pair of CASSCF/auc-cc-pV$X$Z calculations. Following~\cite{VAR19:8022}, $\beta$ is set to 2.284 and a reherarquization procedure has been employed which converts $X=Q,5$ into $x=4.41,5.34 $ to allow for an accurate dual-level CBS  extrapolation of the CASSCF energies. 

The dynamical correlation ($dc$) energy is CBS extrapolated  according to the USTE method~\citep{VAR07:244105},
in which the $dc$ energy of the $X=Q,5$ pair of basis sets are fitted to
\begin{equation}
E_X^{dc}=E_{\infty}^{dc}+\frac{A_3}{(X-3/8)^3}+\frac{A_5^{(0)}+cA_3^n}{(X-3/8)^5}
\label{cbs2}
\end{equation}
which provides $E_{\infty}^{dc}$ and $A_3$. The parameters $A_5^{(0)}$, $c$ and $n$ are system-independent and constant for a given post-HF method~\citep{VAR07:244105}. Their values for MRCI calculations 
are $A_5^{(0)}$=0.0037685459 $E_h$, $c$=-1.17847713 $E_h^{1-(n)}$ and $n$=1.25~\citep{VAR07:244105}.

\subsection{Potential energy surface modeling}

The CBS extrapolated {\em ab initio} points are then fitted to an analytic function using
the DMBE formalism.~\citep{VAR85:401,VAR88:255,VAR92:333,VAR00:33} Accordingly, the PES assumes the form
\begin{equation}
V({ R})=\sum_{i=1}^{3}V_i^{(2)}(R_i)+V^{(3)}(R_{\rm SiS},R_{\rm SiH},R_{\rm SH})
\label{eq:VDMBE}
\end{equation}
where $V_i^{(2)}(R_i)$ are two-body energy terms, and $V^{(3)}$ a three-body one. Both are further partitioned to describe separately short-range ($V_{\rm EHF}^{(n)}$) and long-range ($V_{\rm dc}^{(n)}$) interactions. The latter 
play a major role on barrier-free reactions, specially at low temperatures, which makes them a very important aspect for the present study.
Specifically, we incorporate electrostatic ($R^{-4}$ and $R^{-5}$) and dispersion energy terms ($R^{-6}$, $R^{-8}$ and $R^{-10}$) suitably modeled from accurately calculated dipolar and quadrupolar moments as well as polarizabilities. 

After the long-range interactions are modeled, the \emph{ab initio} energies are fitted to  the $V_{\rm EHF}^{(n)}$ terms such that the final PES (V({ R}))
accurately represents the CBS energies. For this, a polynomial times a range decaying function is employed.
It should be noted that the \emph{ab initio} calculations show the
existence of a conical intersection between $1^2A'$ and $2^2A'$ electronic states for linear SiS-H configurations, 
which causes a cusp in the ground adiabatic PES that is being modeled. Because such cusps cannot be modeled with smooth functions, we improve their description by using a method that allows for a realistic description 
of conical intersections (CI) on adiabatic PESs~\citep{GAL15:1415,GAL16:55}; 
for further details, the reader is addressed to previous work~\citep{GAL15:1415,GAL16:55,ROC016:064309,GON18:4198} where the method has been successfully applied in a variety of triatomic systems.

\subsection{Rate coefficients}

To compute the rate coefficients for the title reaction, we employed the QCT~\citep{VENUS96,PES99:171} method 
using a version of the Venus96~\citep{VENUS96} code costumized by the Coimbra Theoretical and Computational Group. 
 For each temperature, a batch of $10^4$ trajectories were integrated, { starting with reactants separated by 25\,\AA \,and} 
with the initial conditions of every trajectory sampled such that the whole batch mimics the correct 
energy distributions at the fixed temperature. In this spirit, the atom-diatom translational energy
follows the Maxwell-Boltzmann distribution, the rovibrational quantum states of SH follow 
Boltzmann's distribution, and the impact parameter ($b$) is uniformly distributed from 0 to $b_{max}$, 
{ where $b_{max}$ was independently tuned for each temperature, ranging from 20\AA \, for 10K to 9\AA \, for 1000K.
A time step of $0.2\,\rm fs$ was used to ensure a satisfactory conservation of the total energy in all trajectories,
with an average error of $5\times 10^{-4}\,\rm kJ\, mol^{-1}$, with the worst trajectory among all 110000 deviating by only $1.8\,\rm kJ\, mol^{-1}$}.

The Monte-Carlo integrated rate coefficient for the reaction assumes the form
\begin{equation}
 k(T)=g_e(T)\left(\frac{8k_BT}{\pi \mu}\right)^{1/2} \pi b_{max}^2\frac{N^{r}}{N}
\label{eq:kt}
\end{equation}
where $k_B$ is the Boltzmann constant, $\mu$ the reduced mass of the
reactants, and $N^{r}/N$ is the fraction of reactive trajectories. In turn,
$g_e(T)$ is the electronic degeneracy factor, which for the ${\rm Si}(^3P)+{\rm SH}(X^2\Pi)$ collisions assumes the form
\begin {equation}
\begin{split}
 &g_e(T)=2\times\left[2+2\exp{(-542.36/T)} \right]^{-1} \times  \\
 &\left[1+3\exp{(-110.95/T)}+5\exp{(-321.07/T)}\right]^{-1} 
\label{eq:edeg}
\end{split}
\end {equation}
{ To estimate an error margin for the rate coefficients caused by the trajectory sampling, we plot $k(T)$ with error bars corresponding to the statistical 68\,\% confidence interval, which} is given by $\Delta k=k \left(\frac{N-N^{r}}{NN^{r}}\right)^{1/2}$.

\subsection{Astrochemical modeling}

To study the impact of the new proposed reaction for intertellar chemistry, we have used the Nautilus gas-grain code \citep{Ruaud2016} and simulated a shock similar to L1157-B1 ~\citep{podio2017silicon}. Time dependent models with and without the newly proposed reaction were performed to assess its impact  on the SiS abundances.

Nautilus is a full gas-grain astrochemical model that can compute the ice and gas chemical composition under interstellar conditions. The gas-phase chemistry is based on the kida.uva.2014 network \citep{2015ApJS..217...20W}. In addition for all appropriate gas-phase processes, the model computes the adsorption and desorption of the species onto and from the surfaces as well as the reactions on the surfaces. The model is a 3-phase model, which means that in addition to the gas-phase, the species on top of the grains are differentiated as "surface" or "bulk". The surface represents the most external two layers of molecules while the rest below is considered as bulk. Only species from the surface can desorb (through thermal desorption, chemical desorption, cosmic-ray induced desorption, and photo-desorption). Diffusion of the species is faster than in the bulk but both are allowed. Photo-dissociations (due to direct or indirect photons) can occur both on the surface and in the bulk. Only "typical" silicate grains (of 0.1 micron in radius) are considered.  All chemical parameters are the same as in \citep{Ruaud2016}. We refer to this paper for a full description of the code. 

Since the SiS molecule has been observed in shock regions, we have used a three step method. As a first step, we simulated a cold molecular cloud chemical composition by computing the chemical evolution for $10^5$~yr, starting from an initial atomic abundance except for hydrogen, which is entirely molecular. The list of elemental abundances is given in Table \ref{tab:example_table} and are the atomic abundances observed in the $\zeta$ Oph diffuse medium from \cite{2009ApJ...700.1299J} \citep[see also][for discussion]{2018A&A...611A..96R}. The gas and dust temperatures were set to 10K and the total H density to $2\times 10^4$~cm$^{-3}$. The integration time is reasonable for the molecular cloud “chemical age” derived from the comparing observations and modeling \citep[see for instance][]{2013ChRv..113.8710A}. 
The value we have used for the silicon elemental abundance is rather high with respect to the abundance very often considered in the "low metal" case \citep{1982ApJS...48..321G} but it is the one observed in the diffuse medium and that can be reasonaly assumed to be available for volatile interstellar chemistry. The sulphur abundance is also high but the recent chemical models have been able to reproduce cold core observations with such high elemental abundance of sulphur \citep{2017MNRAS.469..435V}. 
The second step of our model simulates the shock itself. Starting from the chemical composition obtained at the end of step 1, we then set the temperature to 100 K, keeping the density the same. 
We run the model for $10^3$yr, 
with the final ice and gas compositions then used as an initial composition for the third step, during which the gas and dust cools down to 30 K and the density then increases to $10^{7}$ cm$^{-3}$. These physical conditions are in agreement with those observed in sources like L1157-B1, which is a protostellar shock driven by the low-mass Class 0 source L1157-mm. 

Several S- and Si-bearing molecules have been catalogued and analysed towards the L1157-B1 shock region (e.g. \citealt{Bachiller1997,Lefloch2012,podio2017silicon,Holdship2019,Feng2020,Spezzano2020}); regarding SiS, \citet{podio2017silicon} reported its detection based on single-dish and interferometric observations towards the shock. They observed various lines of SiS and used the rotational diagram method to obtain an estimate of the gas temperature and SiS average column densities over the cavity of L1157-B1. Their results suggest a gas temperature around 24~K and an abundance value of [SiS/H$_2$] $\sim$ 2 $\times$ 10$^{-8}$ for the head of the cavity. In comparison with SiO, they also found that SiO/SiS ratios might vary in the shock, estimating values of i.e. 25 and over 180 in the head of the cavity B1c and at the jet impact region B1a, respectively. 

We did two sets of models. In the first, we have used the original chemical networks while, in the second, we have included the new reaction $\rm Si + HS \rightarrow SiS + H$.

\begin{table}
	\centering
	\caption{Elemental abundances used for the model \citep{2009ApJ...700.1299J,2018A&A...611A..96R}. }
	\label{tab:example_table}
	\begin{tabular}{ll} 
		\hline
		Species & Abundances \\
		\hline
		 H      &  1.0 \\
		 He     &  0.09 \\
		 N      &  6.2 $\times$ 10$^{-5}$ \\
		 O      &  3.3 $\times$ 10$^{-4}$ \\
		 C$^+$  &  1.8 $\times$ 10$^{-4}$ \\
		 S$^+$  &  1.5 $\times$ 10$^{-5}$ \\
		 Si$^+$ &  1.8 $\times$ 10$^{-6}$ \\
		 Fe$^+$ &  2.0 $\times$ 10$^{-7}$ \\
		 Na$^+$ &  2.3 $\times$ 10$^{-7}$ \\
		 Mg$^+$ &  2.3 $\times$ 10$^{-6}$ \\
		 P$^+$  &  7.8 $\times$ 10$^{-8}$ \\
		 Cl$^+$ &  3.4 $\times$ 10$^{-8}$ \\
		 F      &  1.8 $\times$ 10$^{-8}$ \\
		\hline
	\end{tabular}
\end{table}

\section{Results}

\subsection{Features of the modeled PES}

The final fit employed in the construction of the PES used 212 linear parameters and 56 nonlinear ones, which
mimicked all 2267 energy points with an overall root-mean-square deviation below $\rm 1\, kcal\,mol^{-1}$.
As first pointed out by~\cite{ROS18:695}, this triatomic system has two covalently bound potential energy minima.
The deepest one corresponds to the thiosilaformyl (HSiS) radical, which can  isomerize
via a hydrogen atom migration to the SiSH isomer. Both structures present a nonlinear geometry. The isomerization
transition state (TS1) lies lower in energy than any dissociation channel. 
In collisions between Si+SH, the highest enery isomer (SiSH) is formed in a direct manner, while the other one can be
accessed after the isomerization step. { 
The $\rm S+SiH$ asymptote lies 55 kJ\,mol$^{-1}$ higher in energy than Si+SH, 
hence collisions leading to this channel unlikely contribute
to the abundance of silylidyne (SiH) in the ISM, except at very high temperature regions.}

\begin{figure}\centering
\includegraphics[width=0.55\textwidth]{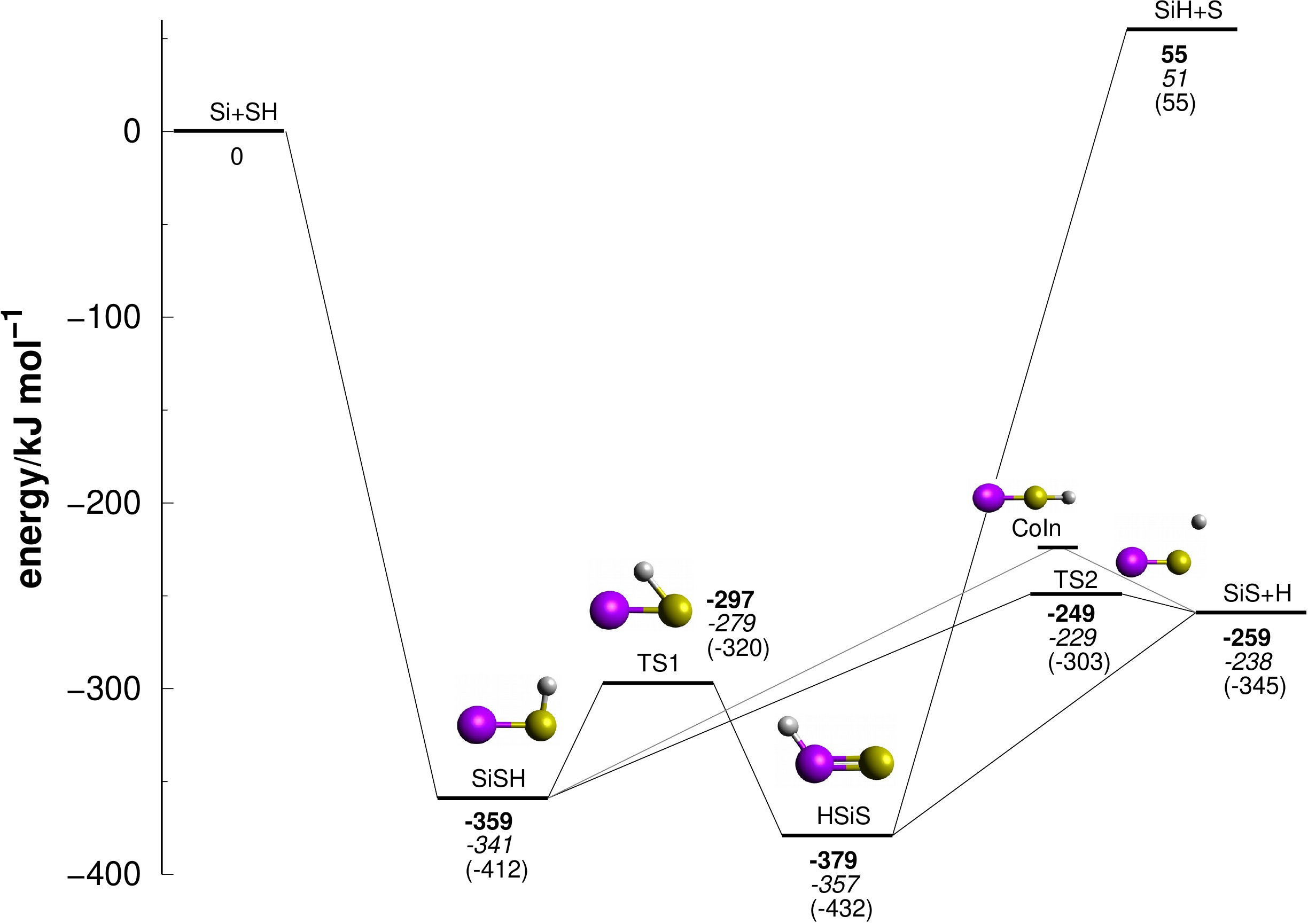}
{\caption{\small \baselineskip=3.0ex 
Schematic representation of the potential energy surface for the reaction, including ZPE correction. { The bold numbers refer to the values obtained by the fit performed in this work, while those of~\cite{ROS19:306} and~\cite{DEL20:175203} are in italic and within parenthesis, respectively.
The conical intersection (CoIn) is also presented and connected with gray lines, to emphasize that this is not a minimum energy path}. Atoms are colored as
follows: silicon (purple); sulfur (yellow); hydrogen (white).}
  \label{fig:PES}}
\end{figure}

{ Figure~\ref{fig:PES} summarizes the main attributes of the potential energy surface. The zero-point energy (ZPE) corrected
values for the stationary structures obtained by the present PES are given in bold. For comparison, also shown are the values obtained by~\cite{ROS19:306} at the CCSD(T)/aug-cc-pV(T+d)Z level and those of~\cite{DEL20:175203}. As seen, 
our results agree well with the CCSD(T) ones. Since our energies are at the CBS limit, and the HSiS minimum shows a relatively high multireference character (T1=0.035,~\cite{GOE21:3647}), we trust our results as the most accurate. 
The agreement with the results of~\cite{DEL20:175203} is acceptable only at the SiH+S asymptotic channel.
}

\begin{figure}\centering
\includegraphics[width=0.55\textwidth]{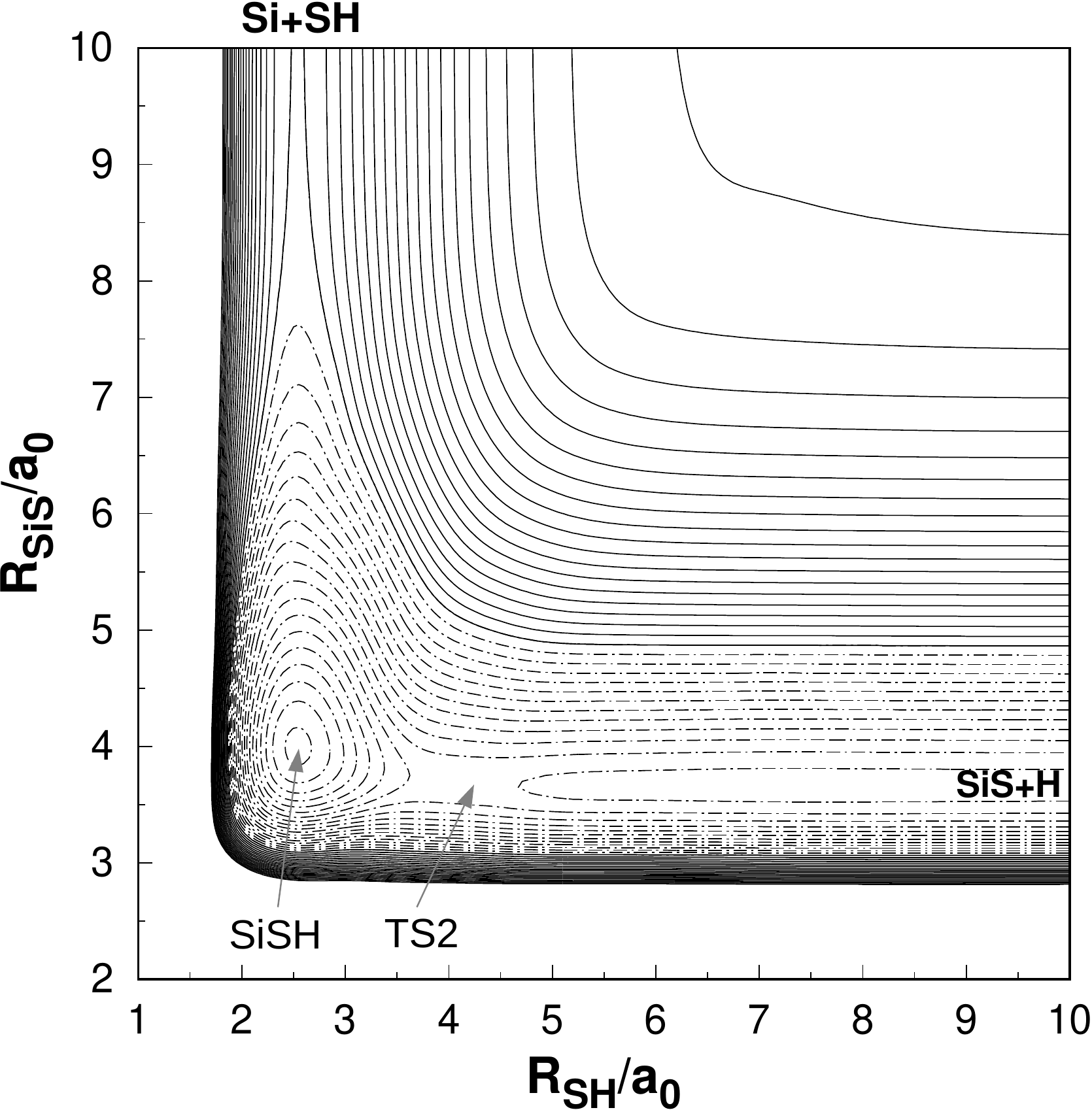}
{\caption{\small \baselineskip=3.0ex Contour plot for a Si-S-H bond stretching at fixed valence angle of 97.3$^\circ$ (showing the SiSH minimum). The zero of energy is set at the reactants limit (Si+SH).
Solid contours correspond to energies above this limit, while dash-dotted ones are lower. The contours are incremented by $\pm 20$ kJ\,mol$^{-1}$.}
  \label{fig:stretch}}
\end{figure}

Fig.~\ref{fig:stretch} provides an overview of the abstraction route towards the products as $\rm Si + SH \rightarrow SiSH \rightarrow SiS + H$.
It shows a contour plot for a reaction occuring at a fixed SiSH angle of 97.3$^\circ$, where it can be seen that the interaction between atomic silicon and the mercapto radical leads to the SiSH minimum without a potential energy barrier.
From this minimum to SiS+H dissociation, there is a small exit transition state (TS2). This barrier lies much lower in energy
than the Si+SH reactants, and hence does not prevent reaction to occur at low temperatures, as it will be shown later.

\begin{figure}\centering
\includegraphics[width=0.55\textwidth]{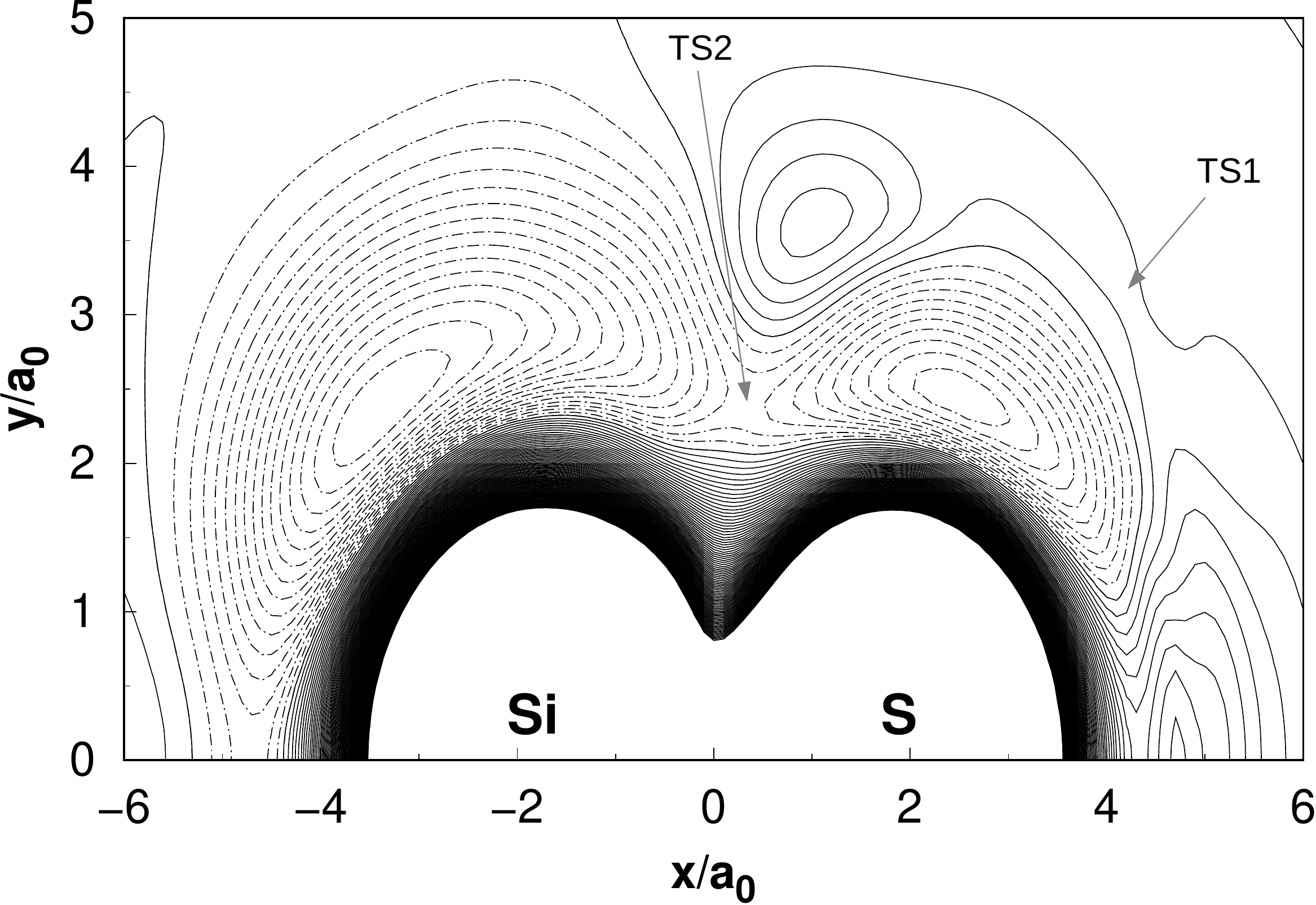}
{\caption{\small \baselineskip=3.0ex 
Contour plots for H  moving around SiS. The zero of energy is set at the products limit (H+SiS).
Solid contours correspond to energies above this limit, while dashed ones are lower. The contours are incremented by $\pm 8.5$ kJ\,mol$^{-1}$.}
  \label{fig:Haround}}
\end{figure}

For an improved visualisation of the two isomers of the triatomic molecule, Fig.~\ref{fig:Haround} shows a contour plot for a hydrogen atom moving around the SiS diatomic with its bond length fixed at equilibrium position. The SiSH isomer that can be obtained directly from reactants is seen in the right hand side of this plot, and TS2 can also be seen above the SiS+H products, as mentioned earlier. 
{ The conical intersection (CoIn) that was modeled by our fitted PES can also be seen in this figure (see also Fig.~\ref{fig:PES}). It occurs when going from the SiSH potential energy basin towards SiS+H but for linear arrangements, which is not the minimum energy path for reaction. The cusp caused by this conical intersection is seen as the sharp contours on the lower part of right hand side of Fig.~\ref{fig:Haround}.
}

{ Fig.~\ref{fig:Haround} also shows the possible H atom migration between the two isomers (through TS1)}. Note that the isomerizing mechanism $\rm Si + SH \rightarrow SiSH \rightarrow HSiS \rightarrow SiS + H$ does not show an exit transition state. This means that the inverse reaction $\rm SiS + H \rightarrow HSiS$ can occur barrierlesly. Although { the SiH+S or Si+SH channels cannot be accessed from SiS+H collisions, the barrierless path to the HSiS minimum} should have important consequences to the SiS collisional vibrational energy transfer. This subject will be investigated in the future.

\subsection{Quasiclassical calculations}

\begin{figure}\centering
\includegraphics[width=0.55\textwidth]{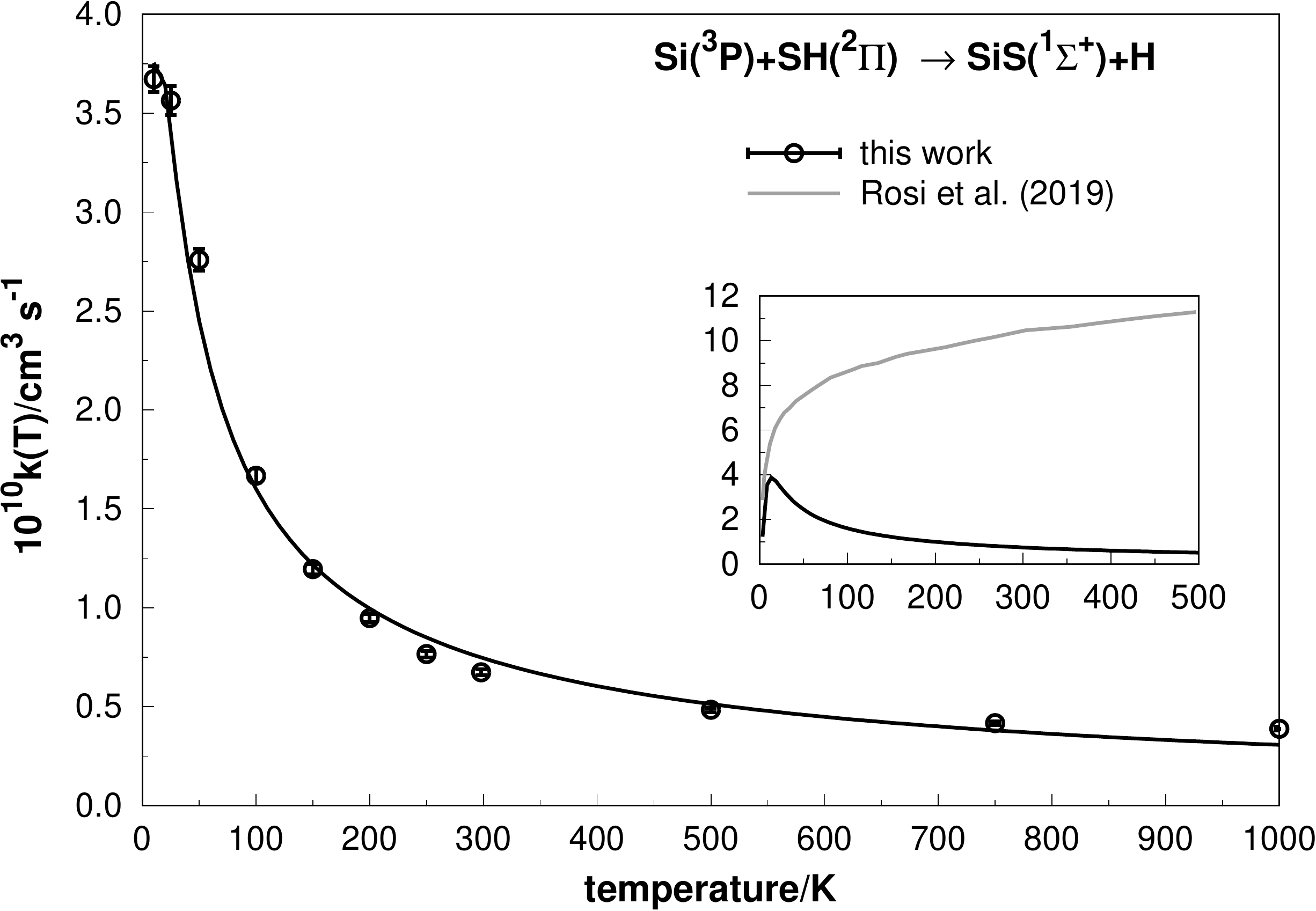}
{\caption{\small \baselineskip=3.0ex Calculated rate coefficients for the Si+SH reaction. { The errorbars correspond to a statistical 68\,\% confidence interval, and for most temperatures are smaller than the circles. The results of ~\cite{ROS19:306} using a capture model are shown in the inset for comparison.} }
  \label{fig:ratec}}
\end{figure}

After integrating $10^4$ trajectories for each temperature and using them to calculate the rate coefficients, the final results are displayed in Fig.~\ref{fig:ratec}.
As shown, the calculated rate coefficients below $\rm 200\,K$ increase drastically with decreasing temperature, as expected for a barrierless reaction. 
In fact, they are similar to the ones reported by \cite{SAN14:335} for the analogous $\rm Si+OH \rightarrow SiO+H$ reaction. 
Moreover, they can be fitted to the modified Arrhenius expression:
\begin {equation}
 k(T)=\alpha \left( \frac{T}{300} \right)^{\beta} \exp{(-\gamma/T)}
\label{eq:Arrh}
\end {equation}
where
$\alpha\!=\!0.770\times 10^{-10}\rm cm^3\,s^{-1}$, 
$\beta\!=\!-0.756$ and 
$\gamma\!=\!9.873\, \rm K$.
%
%
%

We should point out at this stage that previous attempts to estimate the rate coefficients for 
the Si+SH reaction have been proposed~\citep{ROS19:306} within a simple capture model, achieving a high temperature limit of 
$1\times 10^{-9}\rm cm^3\, s^{-1}$. { This is higher than our results and is also shown for comparison in Fig.~\ref{fig:ratec}}. 
The capture model does not show the same temperature dependence, { with rate coefficients decreasing for lower temperatures}.
Another QCT study has been performed~\citep{DEL20:175203}
but the authors predict that the reaction shows a barrier-like behavior, { thus contradicting} other theoretical work~\citep{ROS18:695,ROS19:306,PAI20:299}
and the possibility of using a capture-type model since this implies a barrier-free reaction. Not surprisingly, their thermal rate
coefficients are much smaller than ours and { nearly} four orders of  magnitude smaller then that of the analogous Si+OH reaction. In the absence of evidence that justifies
the existence of any barrier, we trust the present work
as realistic and appropriate for astrochemical databases.


\subsection{Astrochemical modeling results}

\begin{figure}\centering
	\includegraphics[width=0.55\textwidth]{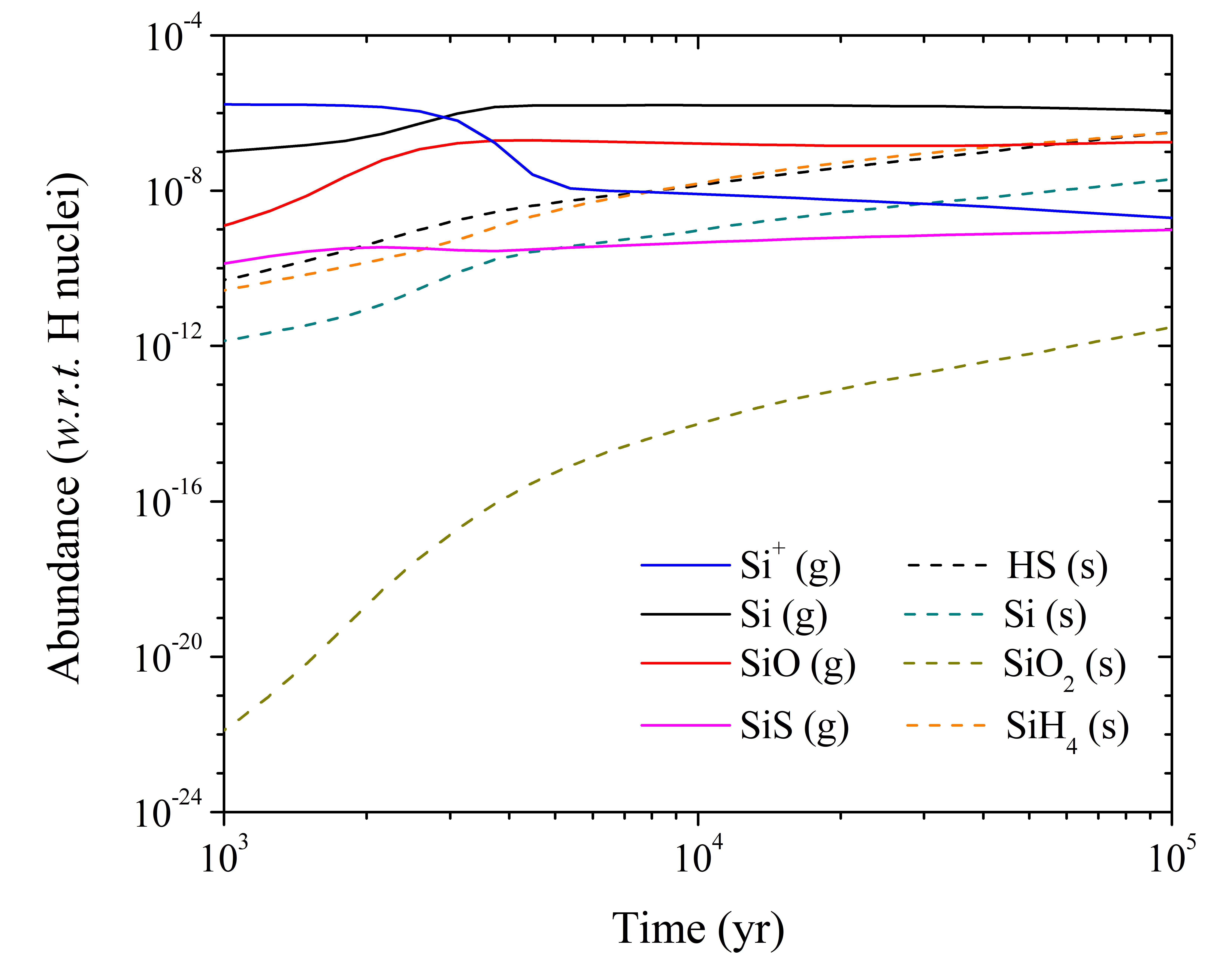}
    \caption{Abundances as a function of time obtained from the step corresponding to the cold core phase. The terms in parentheses, ``g'' and ``s''  \ stand for abundances computed in the gaseous and solid phase, respectively.\label{fig1}}
    
\end{figure}

During the first step (cold molecular chemistry), given in Fig.~\ref{fig1}, the silicon, initially in its atomic ionised form, is mostly neutralised and the neutral Si in the gas-phase is still very high at $10^5$ yr. Part of the silicon depletes on the dust surfaces, forming SiO and mostly SiH$_4$. 
Sulphur is also first neutralised and then depletes forming almost equally HS and H$_2$S on the grains \citep[see also][]{2017MNRAS.469..435V}. 

\begin{figure}\centering
	\includegraphics[width=0.55\textwidth]{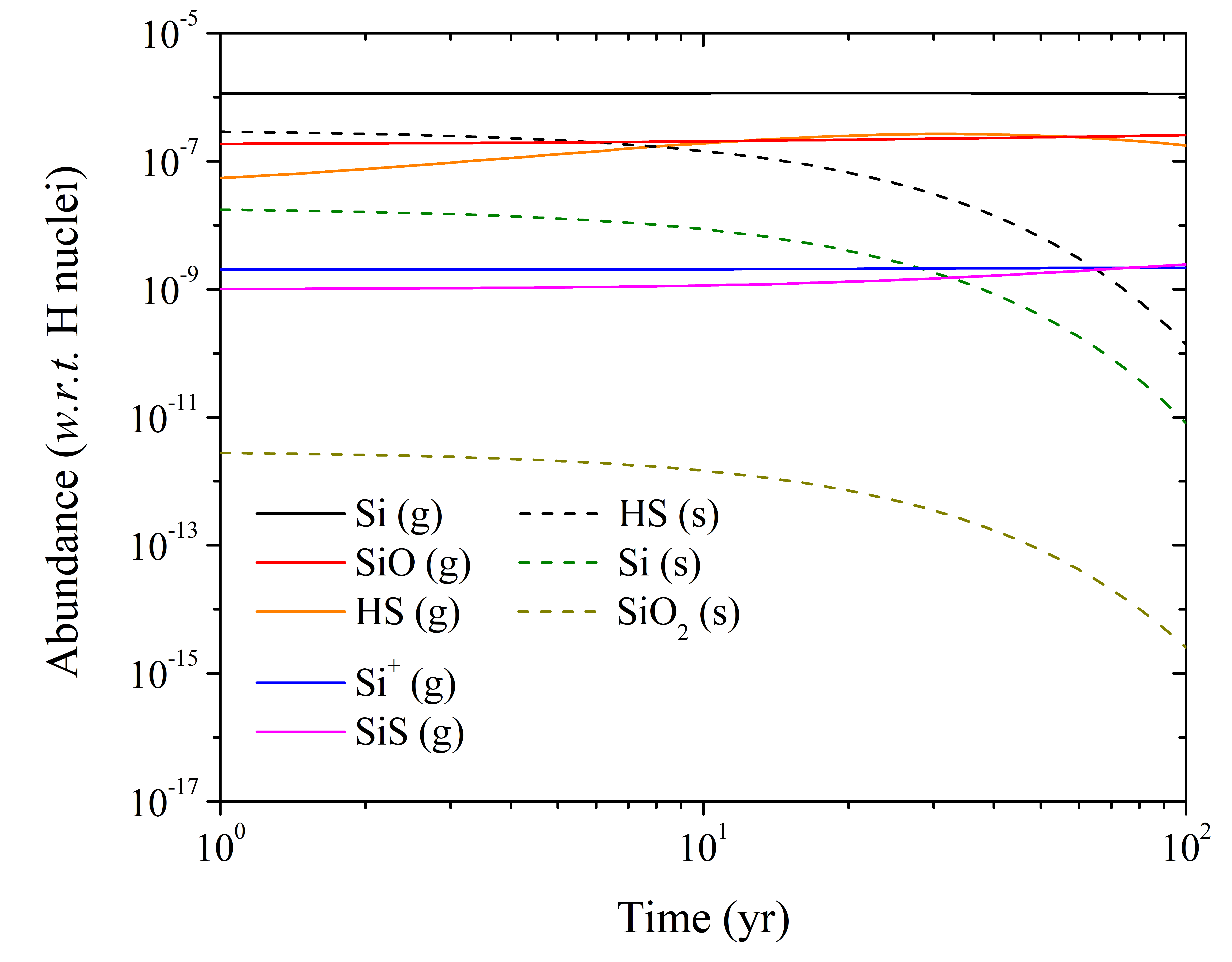}
    \caption{Abundances as a function of time obtained from the second and fast step. As indicated in Fig.~\ref{fig1}, the terms  ``g'' and ``s''  stand for abundances computed in the gaseous and solid phase, respectively.}
    \label{fig2}
\end{figure}

The results of the second fast step can be seen in Fig.~\ref{fig2}, where the only effect of temperature is to desorb the Si and S-molecules in the gas-phase. The Si gas-phase atomic abundance already high at the beginning of this step stays as high as {$\sim 2 \times 10^{-6}$} (Fig.~\ref{fig2}).
\ HS being desorbed from the grains produces an abundance to more than 1 $\times$ 10$^{-7}$ (see Fig.~\ref{fig2}). 


\begin{figure*}\centering
	\includegraphics[width=0.55\textwidth]{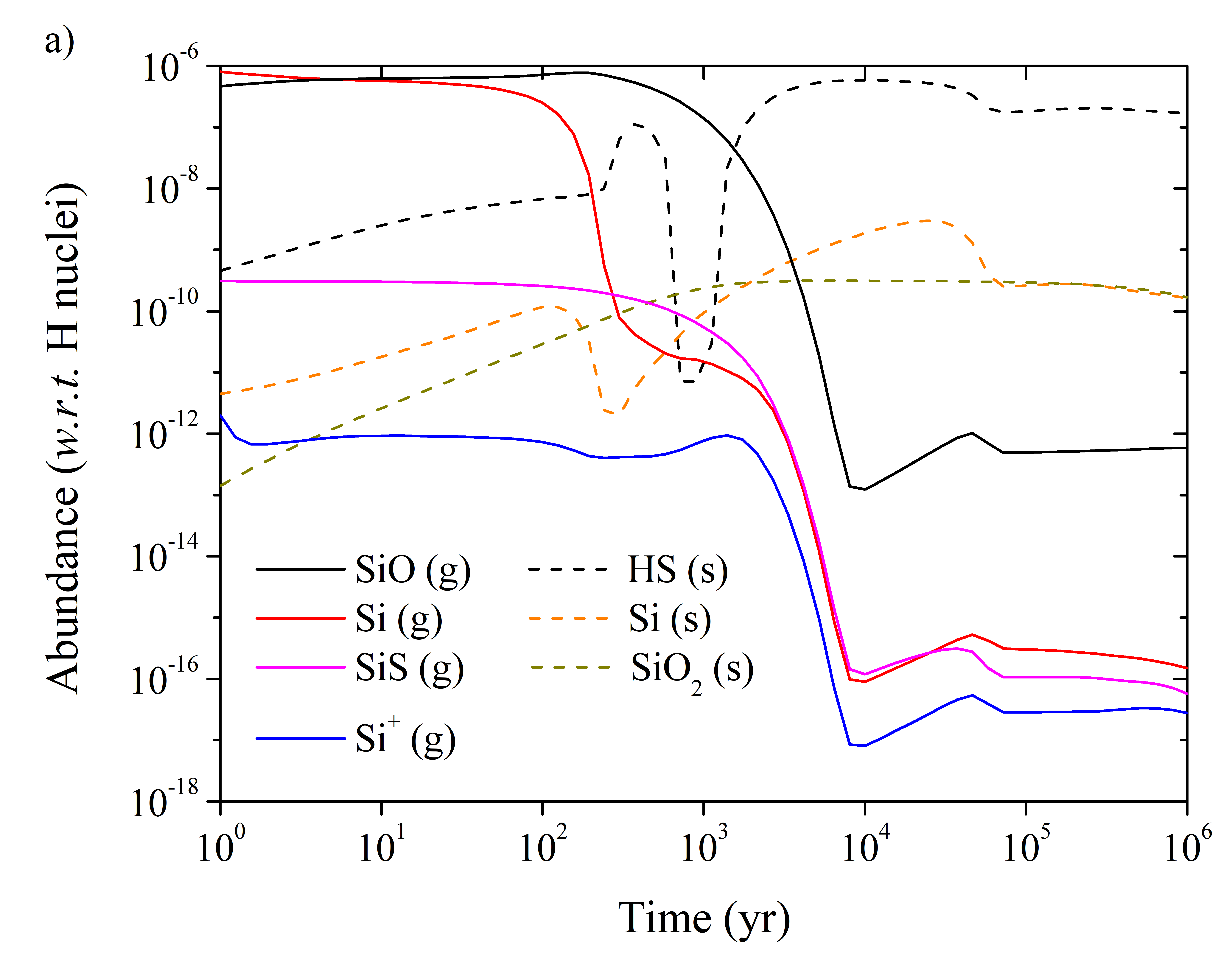}
	\includegraphics[width=0.55\textwidth]{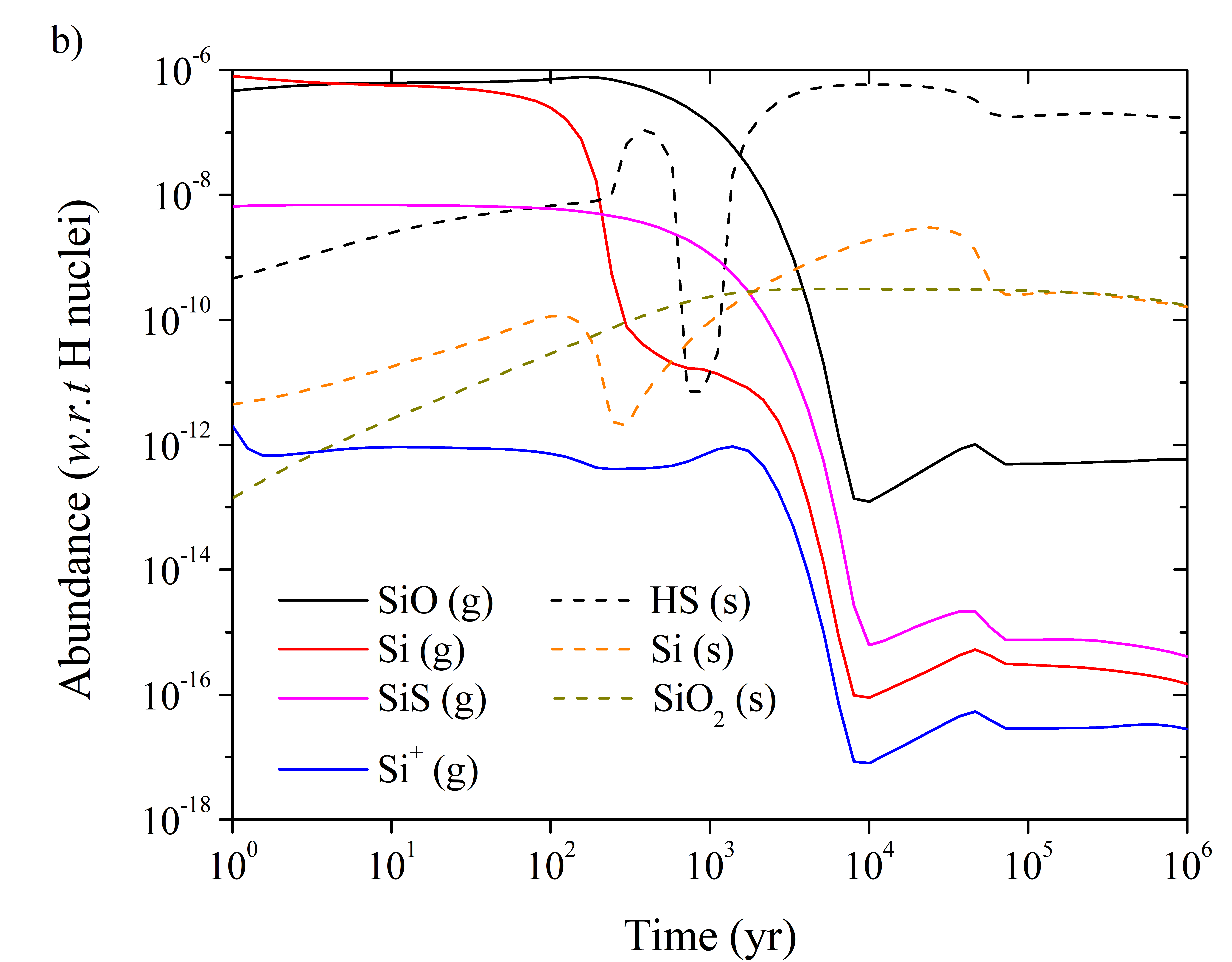}
    \caption{Abundances as a function of time obtained from the step corresponding to the shock model: comparative results a) without and b) with the reaction Si + HS $\rightarrow$ SiS + H. Similar to Fig.~\ref{fig1}, the terms  ``g'' and ``s''  stand for abundances computed in the gaseous and solid phase, respectively. }
    \label{fig3}
\end{figure*}

Starting the last step (step 3) of the model  with such high Si and SH abundances in the gas-phase, and with the new neutral-neutral reaction included in the model, strongly enhanced the SiS abundance predicted by the model, as can be seen in Fig.~\ref{fig3}. 
Such increment is, for instance, $\sim$~28 times at the early ages ($< 10^2$~yr) of the shock model, with abundance values (relative to atomic hydrogen) rising from $\sim$~2.5 $\times$ 10$^{-10}$ to 7 $\times$ 10$^{-9}$. After this time, the silicon depletes again on the grains. The new abundance relative to molecular hydrogen is therefore $1.4\times 10^{-8}$, which matches very well the observed abundance of [SiS/H$_2$]$\sim 2\times 10^{-8}$~\citep{podio2017silicon}.

{ To assess the dependency of the results with changes in the rate coefficients, we also employ the values obtained by ~\cite{ROS19:306}
to rerun the same astrochemical model. These rate coefficients are about one order of magnitude higher than ours, and approximate our values only below 25K. 
As a result, the time evolution of the abundances stay essentially the same as those previously presented in Figures ~\ref{fig1},~\ref{fig2} and~\ref{fig3}. 
However, the final abundance of SiS relative to molecular hydrogen is approximately 2.5 higher than when using our rate coefficients, attaining a value of about  $3.5\times 10^{-8}$. 
In summary, using the rate coefficients of~\cite{ROS19:306} yields an abundance higher than the observed one, while deviating from it more than when using our calculated values. Nevertheless, the predicted abundances of SiS compare reasonably well with the observed values within a given observational uncertainty.
}

\section{Concluding remarks}
In this work we have performed accurate multireference calculations based on the MRCI(Q) method, which were subsequently extrapolated to the complete basis set limit. The calculated energies cover the whole configurational space of the HSiS triatomic and were accurately modeled via DMBE method. This approach further warrants an accurate representation of the long-range interactions, which are of crucial importance for predicting rate coefficients at low temperatures. The fitted PES was then employed in QCT calculations of the rate coefficients for the Si+SH$\rightarrow$SiS+H reaction over a wide range of temperatures.

The relevance of the title reaction for SiS formation was tested with an astrochemical gas-grain model of the  L1157-B1 shock region. The results show that the inclusion of this neutral-neutral process enhances by 28-fold the SiS gas-phase abundance relative to \ce{H2}, thence from $5\times 10^{-10}$ to $1.4\times 10^{-8}$. It is therefore sufficient to explain the observed abundance of [SiS/H$_2$]$\sim 2\times 10^{-8}$. 
This reaction helps therefore on elucidating the mechanisms of formation and abundance of silicon monosulfide in the ISM. 

The new rate coefficients can be used in other astrochemical models and also be added to existing databases. Aditionally, the novel potential energy surface can be used to predict the rate coefficients for other molecular collisions such as S+SiH and SiS+H, which are also relevant for astrochemistry.
{ Additional details regarding the potential energy surface, comparison with previous work, treatment of the ZPE leakage (\cite{VAR07:386}), and references therein) in QCT trajectories and state specific rate coefficients will be given in a posterior publication.}

{\allowdisplaybreaks
\section*{Acknowledgments} 

This work has been financed in part by the Coordena\c c\~ao de Aperfei\c coamento de Pessoal de N\'ivel Superior - Brasil (CAPES) - Finance Code 001,
Conselho Nacional de Desenvolvimento Cient\'ifico e Tecnol\'ogico (CNPq), grant 305469/2018-5 (BRLG) and grant 150465/2019-0 (E.M.), and
Funda\c c\~ao de Amparo \`a Pesquisa do estado de Minas Gerais (FAPEMIG). The authors also acknowledge the CNRS program ``Physique et Chimie du Milieu Interstellaire'' (PCMI) co-funded by the Centre National d’Etudes Spatiales (CNES).
BRLG is also thankful to Rede Mineira de Química (RQ-MG), while VCM acknowledges
the support of Edital 2015 do Programa institucional de Fundo de Apoio \`a Pesquisa da Universidade Federal do Esp\'irito
Santo and also the Edital Universal FAPES 2018.
This work has also the support of Foundation for Science and Technology, Portugal, and Coimbra Chemistry Centre,
Portugal, through the project UID/QUI/00313/2019. AJCV thanks also
China’s Shandong Province ``Double-Hundred Talent Plan'' (2018).}




\end{document}